# A FIXED POINT IN COPTIC CHRONOLOGY: THE SOLAR ECLIPSE OF 10 MARCH, 601

At the very beginning of his article in the *Coptic Encyclopedia* on the dating of inscriptions, Pierre Du Bourguet remarked, "Broadly speaking, the dating of Coptic monuments and artifacts is one of the thorniest problems in this archaeological field".[1]  The purpose of this paper is to reconsider one of the rare exceptions to this gloomy generalisation, a fixed point in the Coptic record - and in Coptic palaeography - which has received surprisingly little attention in the years since it was first published, well over a century ago.

In 1878, Ludwig Stern published a short note on a Coptic ostracon in the Egyptian Museum in Turin, which gives a clear reference to an eclipse of the sun.[2]  This text runs as follows:

| † | hN soy mNtafte M | | ayw hN terompe |
|---|---|---|---|
| | PamenwQ tEs | | etere petros M |
| | tetartEs indik\ | | paloy o NlaSane |
| | a prE Rkake NXp | | eXEme NhEts |
| | fto Mpehooy   † | | † |

"On the fourteenth day of Phamenoth of the fourth (year of the) indiction, the sun went dark at the fourth hour of the day - and in the year when Petros, son of Palou was magistrate over Djême."

The text itself is straightforward, and its meaning was clear to the original editor. The inscription records an eclipse of the sun on a specific day at as specific time and in a particular place, Djême (Medinet Habu) on the west bank at Thebes.  The problem lies in identifying the year, since this account, like many other Coptic inscriptions, is dated by an indiction-cycle instead of any absolute scale.  The indiction cycle covers a fifteen-year period, whereupon a new cycle repeats itself. This system is adequate for the use to which it was originally put, namely calculating the payment of taxation and keeping short-term records, but it is obvious that it cannot provide a more precise date.  This may well be the reason why the second part of the text adds the name of a particular magistrate, who would act as an eponym for the year or years when he held office. However, our knowledge of the magistrates at Djême is nowhere near complete, and the mere mention of a name does nothing to identify the year in question. Ludwig Stern was therefore obliged to fall back on the other records which have survived from the site of Djême, which he was inclined to date to the late seventh or the eighth centuries AD, the period to which many of the papyri found at the site can be assigned.[3]

This way of dating can only be approximate, as Stern readily admitted. Fortunately, modern astronomical knowledge enables historical eclipses to be identified with great accuracy, and the text from Djême turns out to be no exception. By far the best candidate for the Djême eclipse is the solar one of 10 March, AD 601. This was first suggested before 1902 by the astronomers H. H. Turner and J. K. Fotheringham, the latter of whom devoted much of his career to ancient astronomy and its implications for chronology.[4] More detailed information on historical eclipses is now available to confirm this suggestion, and to supplement it.

The eclipse of 10 March 601 was the only total eclipse visible over the lower Nile valley in the period from 570 until 738. It was total at the latitude of modern Cairo, just north of Memphis, and the totality lasted for 2 minutes 46 seconds, which is substantial for a solar eclipse. At Cairo (to continue a convenient anachronism), the maximum occurred at 09:55:57 in the morning, local time. Totality extended roughly to the latitude of present-day Minya. At the latitude of Djême the eclipse was not quite total; however, with a coverage of 93% the event would still have been very noticeable. Maximum at Luxor fell slightly later in the morning than at Memphis, at 09:57:13. Local sunrise was at a few minutes after 06:00 (depending on the conditions of the horizon), so that this agrees perfectly with the reference to the fourth hour of the day in the Turin ostracon. The day in question, 10 March, similarly corresponds to the 14 Phamenoth of the Coptic text. The year is also compatible with what we know of the 15-year indiction-cycle. This practice was instituted in AD 312/3, and a new cycle will have begun in 597/8. The fourth year of this cycle was 600/601. On the same day, 10 March, in AD 582 there was a partial eclipse of the sun visible in Egypt, but this would have been far less impressive, and 582 was not the fourth year of an indiction, but rather the first. The event of 601 therefore remains by far the best candidate for our Coptic eclipse.[5]

It is a commonplace that records of eclipses are almost entirely absent from ancient Egyptian texts.[6] This is in contrast with the state of affairs from Mesopotamia and China, where references to eclipses are well attested. (One of the cuneiform examples, the solar eclipse of 15 June 763 BC, has become one of the essential points of Assyrian chronology.) The reason for the scarcity of eclipses in Egyptian records is not known. One obvious possibility is that eclipses were ill-omened, and therefore not suitable to be recorded in temples or tombs, from where the majority of our evidence comes. The one negative exception which we have, a mention that a catastrophe happened even though there had not been an eclipse to herald it, fully supports the idea that eclipses were dangerous. This is the controversial passage in the 9th-century BC Chronicle of Prince Osorkon which remarks that a convulsion of some kind broke out in the land "even though the sky did not swallow up the moon". The absence of an eclipse could be mentioned in a hieroglyphic text, even if an actual eclipse could not be.[7] Here it may be significant that the eclipse which never happened is envisaged to be a lunar one. Lunar eclipses may have been marginally less unmentionable than solar ones, since they do not normally carry the awesome quality which solar eclipses can convey.

An alternative is that eclipses were written down by the Egyptians, but these accounts were kept in libraries or archives, either in temples or in private hands. Very few collections of this sort have survived from Pharaonic times, and none of these is likely to be complete. However, a work on eclipses in general does survive from the Roman period. This is a demotic papyrus, possibly from the Fayyûm, which derives omens from eclipses both solar and lunar, depending on the hours when they are seen, the months of the year, or particular features of their appearance. The text bears the very strong influence of Mesopotamia, where astrology originated, and it is possible that the underlying scheme of the composition goes back as early as the 6th century BC.[8] This material has been adapted to an Egyptian context, but the borrowing from Mesopotamia is still undeniable. Here we at least have proof that eclipses were observed and reported by the Egyptians in the Roman period, if not earlier, and that studies of such phenomena could be preserved. It may be that the practice was known considerably earlier, but at the moment we have no way of knowing how far back this might have gone.

Some confirmation of the idea that eclipses were important in the Pharaonic period can be found from the one passage which directly mentions an eclipse as taking place. This too is contained in a demotic papyrus, P. Berlin 13588. This was first published by W. Erichsen, and further studied by Mark Smith.[9] The text takes the form of an appeal by a young priest from Daphnae, in the north-east Delta, who lays claim to a share of certain priestly incomes. He strengthens his case by referring to an act of piety which he performed some time previously, and he explains the historical background to this as follows: "I heard in Daphnae, my home town, that the sky had swallowed the disk ... after the sun when it was going to its rest in the evening. The death of Pharaoh occurred in the lands to the east of Na'amepin"as." (P. Berlin 13588, iii, 1-3). As the second editor convincingly showed, the narrator here is describing a lunar eclipse which occurred immediately after sunset, and not a solar eclipse as had previously been thought. This is most likely to have been the lunar eclipse of 22 March 610 BC, which took place around 18:12:0, one or two minutes after sunset, and was visible from Daphnae. The same year saw the death of the Pharaoh Psammetichus I, and in the Berlin text the eclipse and the death are clearly linked, as being two very baleful events. Here too we are dealing with a record of a lunar eclipse rather than a solar one.

The Berlin papyrus is essentially a secular document, rather than a religious record, although the narrative has been elaborated, since it seems to have acquired some kind of literary status. All the same, a mention of an eclipse may well have been acceptable in such a context. The narrator feels justified in referring to an event of this kind, because it serves to show his piety and trustworthiness in a moment of supernatural crisis.

At Alexandria eclipses were certainly observed, and a couple were recorded by Ptolemy in his *Almagest*. However, these were of interest to the astronomer as a means of calibrating the length of the Egyptian year, and they are not related to individual events. They are part of the history of Greek science, rather than Egyptology as generally understood, and they are therefore omitted from this paper.

To return to the Coptic text in Turin. The fact that was written on an ostracon, rather than on the more formal medium of papyrus, may be significant. This modest inscription, written quickly on a potsherd, is an informal record, rather than a liturgical or theological one. It is not part of an epitaph, or a homily, or an encomium

---

of a saint. We do not know why the scribe at Djême made his record, but he clearly felt the need to surround the text with signs of the Cross, in an attempt to invoke divine protection. He may have been uneasy in view of the numerical pattern of an event which took place at the fourth hour on the fourteenth of Phamenoth in the fourth year of an indiction. All this makes our Coptic eclipse all the more valuable, since it is the only example that we have of a datable solar eclipse from the entire length of Egyptian history, before the coming of the Islamic era.

Institute of Astronomy, Cambridge                    Gerry Gilmore

Faculty of Oriental Studies, Cambridge                    John Ray